\documentclass[aps,10pt,notitlepage,twocolumn,nofootinbib,superscriptaddress]{revtex4-2}
\usepackage{graphicx}
\usepackage{amssymb}
\usepackage{amsmath}
\usepackage{subcaption}
\usepackage{placeins}
\usepackage[colorlinks=true,          
            linkcolor=red,          
            citecolor=blue,         
            urlcolor=blue]{hyperref} 
\usepackage{booktabs}
\usepackage{siunitx}
\usepackage{float} 

\begin{document}
\title{Radiation properties of a regular black hole embedded in a Dehnen-type dark matter halo with a thin accretion disk}

\author{Tianyou Ren}
\affiliation{College of Physics Science and Technology, Hebei University, Baoding 071002, China}

\author{Jing-Ya Zhao}
\affiliation{College of Physics Science and Technology, Hebei University, Baoding 071002, China}

\author{Xiaomei Liu}
\affiliation{College of Physics Science and Technology, Hebei University, Baoding 071002, China}

\author{Rong-Jia Yang \footnote{Corresponding author}}
\email{yangrongjia@tsinghua.org.cn}
\affiliation{College of Physics Science and Technology, Hebei University, Baoding 071002, China}
\affiliation{Hebei Key Lab of Optic-Electronic Information and Materials, Hebei University, Baoding 071002, China}
\affiliation{National-Local Joint Engineering Laboratory of New Energy Photoelectric Devices, Hebei University, Baoding 071002, China}
\affiliation{Key Laboratory of High-pricision Computation and Application of Quantum Field Theory of Hebei Province, Hebei University, Baoding 071002, China}

\begin{abstract}
   We investigate the shadow, timelike geodesic structure, radiation properties of thin accretion disks, and optical appearance of a static spherically symmetric regular black hole, constructed based on the Dehnen-type density profile. Using observational data from M87* and Sgr A*, we constrain the model parameter $a$ at both $1\sigma$ and $2\sigma$ confidence levels. Based on the Page-Thorne model, we calculate the local radiative flux, redshift factor distribution, and the radiation flux received by a distant observer, systematically examining the effects of the parameter $a$ and the viewing angle on the black hole image. The results show that larger $a$ will enlarge the effective radiation area of the accretion disk and significantly enhance the asymmetry and Doppler boosting effects of the direct and secondary images at large viewing angles. 
\end{abstract}

\maketitle

\section{Introduction}
\label{introduction}

Black holes are one of the most striking predictions of general relativity, and their strong gravitational field regions provide natural laboratories for testing gravitational theories and exploring extreme physical laws. However, classical black hole solutions inevitably possess curvature singularities inside the event horizon, signaling the breakdown of classical general relativity at extremely high energy scales. The existence of singularities not only leads to geodesic incompleteness of spacetime but also fundamentally conflicts with the unitarity requirements of quantum mechanics, giving rise to the famous black hole information paradox. To address this issue, researchers have proposed various schemes to avoid singularities, including nonlinear electrodynamics, string theory, loop quantum gravity, and gravitational decoupling method, which have led to the important research direction of regular black holes \cite{Lan:2023cvz, Ansoldi:2008jw, Bambi:2023try, Ayon-Beato:1998hmi, Bronnikov:2000vy, Dymnikova:2004zc, Fan:2016hvf, Aydiner:2025eii, Kar:2023dko,  Zhao:2025sck, Saidov:2025mmi, Bambi:2013ufa, Toshmatov:2017zpr, Hua:2025qwu, Hua:2025efb}.

On the other hand, astronomical observations indicate that black holes in the real universe are not isolated but are embedded in galactic environments dominated by dark matter halos \cite{Navarro:1996gj, Bertone:2004pz}. Although the microscopic nature of dark matter remains unknown, $N$-body simulations and galactic dynamics studies have developed several phenomenological density distribution models, such as the Navarro-Frenk-White (NFW) profile \cite{Navarro:1996gj}, the Hernquist profile \cite{Hernquist:1990be}, the Einasto profile \cite{Einasto:1965czb, Haud:1986yj, Frutos-Alfaro:2012gyp}, and the Dehnen-type profile \cite{Dehnen:1993uh}. This opens up the possibility of using dark matter halos both as gravitational sources and as a mechanism for singularity avoidance.

Recently, Konoplya and Zhidenko \cite{Konoplya:2025ect}, based on this idea, systematically constructed a class of static spherically symmetric regular black hole solutions sourced by dark matter halos by imposing the radial pressure condition $P_r = -\rho$. They also demonstrated linear stability against axial perturbations. This model elevates the dark matter halo from a passive environmental background to an active mechanism for resolving singularities, providing a novel theoretical platform for exploring the intersection of strong-field gravity and dark matter physics.

For such black holes embedded in dark matter halos, existing research has extensively investigated observable signals such as quasinormal modes, shadows, lensing, and accretion disk radiation. For instance, Datta \cite{Datta:2023zmd} analyzed in detail the energy conditions for singular black holes within Einasto and Dehnen-type dark matter halos; Konoplya et al. \cite{Konoplya:2022hbl, Konoplya:2025mvj} examined the axial perturbation stability and shadow characteristics of black holes surrounded by dark halos; Anjan and Sayan \cite{Kar:2025phe} systematically reviewed the asymptotic behavior and horizon structure of regular black holes in the Dehnen-type distribution. Furthermore, concerning loop quantum gravity regular black holes (LQRNBH) and quantum-corrected black holes (QCBH), Yu-Heng Shu \cite{Shu:2024tut} used EHT observational data for M87* and Sgr A* to impose high-precision constraints on quantum parameters and systematically studied the radiative efficiency, redshift distribution, and imaging characteristics of thin accretion disks. In other modified gravity frameworks, the radiative properties of thin accretion disks have also been extensively investigated. For example, Liu et al. \cite{Liu:2024brf} revisited the thin accretion disk around brane-world black holes, correcting previous errors in the flux formula and presenting revised results. He et al. \cite{He:2022lrc} analyzed the effects of the model parameters on the energy flux, temperature, and emission spectrum in Einstein-Aether-scalar theory. Gravitational wave counterparts were explored in Ref.~\cite{Yang:2024lmj}. Optical images of the Kerr–Sen black hole were investigated in \cite{Wang:2025buh}. Feng et al. \cite{Feng:2024iqj} systematically calculated the radiative flux, spectrum, and shadow of Kerr-Sen black holes in a plasma environment and constrained the dilaton parameter using M87* and Sgr A* observations. These works indicate that black hole shadow radii and accretion disk radiation spectra are powerful probes for distinguishing different modified gravity models and constraining dark matter halo parameters.

However, for regular black holes embedded in a Dehnen-type dark matter halo (hereafter referred to as Dehnen regular black holes), a systematic study focusing on thin accretion disk imaging is currently lacking. How does the parameter $a$ specifically modulate the circular orbit dynamics, local radiative flux, redshift factor distribution, and the observed radiation image? What patterns of variation emerge from its influence at different viewing angles? What confidence level constraints can be placed on $a$ using existing EHT observations? These questions urgently need answers.

This paper aims to fill this gap. Taking the Dehnen regular black hole metric as the research object, we conduct the following systematic analysis: First, based on EHT observations of M87* and Sgr A* \cite{EventHorizonTelescope:2019dse, EventHorizonTelescope:2019uob, EventHorizonTelescope:2019jan, EventHorizonTelescope:2019ths, EventHorizonTelescope:2019pgp, EventHorizonTelescope:2019ggy, EventHorizonTelescope:2022wkp, EventHorizonTelescope:2022apq, EventHorizonTelescope:2022wok, EventHorizonTelescope:2022exc, EventHorizonTelescope:2022urf, EventHorizonTelescope:2022xqj}, we provide quantitative constraints on the parameter $a$ at both $1\sigma$ and $2\sigma$ confidence levels. Second, we systematically investigate the timelike geodesic structure, revealing the influence of $a$ on the ISCO radius, orbital energy, angular momentum, and angular velocity. Subsequently, employing the Page--Thorne thin accretion disk model \cite{Page:1974he} combined with the backward ray-tracing method \cite{Luminet:1979nyg}, we numerically simulate isoredshift curves and observed radiative flux distributions for different viewing angles, systematically examining the modulating effects of the parameter $a$ on the characteristics of direct and secondary images. Finally, we summarize the multiple observational imprints of the parameter $a$ in black hole shadows, orbital dynamics, and accretion disk imaging, and discuss potential future research directions.

The structure of this paper is organized as follows: Section \ref{BlackHoleMetricAndParameterConstraints} briefly reviews the Dehnen regular black hole metric and provides parameter constraints based on EHT data; Section \ref{MotionOfMassiveTestParticlesAndCircularOrbits} derives timelike geodesics and circular orbit conditions, analyzing the impact of $a$ on the ISCO and orbital dynamics; Section \ref{ObservedFluxRedshiftAndBlackHoleOpticalAppearance} establishes the thin accretion disk model, calculates the redshift factor and observed radiative flux, and presents direct and secondary imaging results for different parameters; Section \ref{Conclusion} summarizes the findings and discusses future prospects.For simplicity, we use natural units ($G=c=1$) throughout this paper.

\section{Black Hole and Parameter Constraints}
\label{BlackHoleMetricAndParameterConstraints}

Finding black hole solutions without singularities is an important research direction in general relativity. This paper is based on a class of regular black hole solutions constructed from the density distribution of dark matter halos, where the dark matter halo density distribution adopts the Dehnen-type profile. This profile has the following form \cite{Dehnen:1993uh, Konoplya:2025ect}:
\begin{equation}
\rho(r) = \rho_0\left(\frac{r}{a}\right)^{-\alpha}\left(1 + \frac{r^k}{a^k}\right)^{-(\gamma -\alpha) / k},
\end{equation}
where $\rho_0$ is the central density, $a$ is the scale parameter, and $\alpha$, $\gamma$, and $k$ are the shape parameters. Taking $\alpha = 0$ to ensure the density is finite at $r=0$ and satisfies the weak energy condition, assuming $\gamma > 3$ to guarantee a finite total mass, and further taking $k = 1$ and $\gamma = 4$, one obtains the following simplified density distribution
\begin{equation}
\rho(r) = \rho_0 \left(1 + \frac{r}{a}\right)^{-4}.
\end{equation}
Using the Einstein field equations and assuming that the radial pressure satisfies $P_r = -\rho$, we can solve to obtain the metric of the following form \cite{Konoplya:2025ect}
\begin{equation}
ds^{2} = -f(r)\text{d}t^{2} + \frac{\text{d}r^{2}}{f(r)} + r^{2}(\text{d}\theta^{2} + \sin^{2}\theta \text{d}\phi^{2}),
\end{equation}
where the metric function $f(r)$ is given by
\begin{equation}
f(r) = 1 - \frac{2M r^2}{(r + a)^3},
\end{equation}
where $M$ is the total mass, related to the central density $\rho_0$ and the scale parameter $a$ by
\begin{equation}
M = \frac{8\pi\rho_0 a^{3}}{6}.
\end{equation}
This metric describes a regular black hole embedded in a Dehnen-type dark matter halo.

Now we study the null geodesics in this spacetime. For photons in the equatorial plane ($\theta = \pi/2$), we have
\begin{equation}
\label{light-like}
\begin{aligned}
0 &= g_{\mu\nu} \left( \frac{\partial}{\partial \lambda} \right)^\mu \left( \frac{\partial}{\partial \lambda} \right)^\nu \\
   & = -f(r) \left( \frac{dt}{d\lambda} \right)^2
    + f(r)^{-1} \left( \frac{dr}{d\lambda} \right)^2
    + r^2 \left( \frac{d\phi}{d\lambda} \right)^2,
    \end{aligned}
\end{equation}
where $\lambda$ is an affine parameter. The metric $g_{\mu\nu}$ does not depend explicitly on the coordinates $t$ and $\phi$, so this spacetime possesses two independent Killing vector fields. For a geodesic in this spacetime, we can derive two conserved quantities, the energy $E$ and the angular momentum $L$
\begin{equation}
\label{E}
E \equiv -g_{00}\frac{dt}{d\lambda}=f(r)\frac{dt}{d\lambda},
\end{equation}
\begin{equation}
\label{L}
L \equiv g_{33}\frac{d\phi}{d\lambda}=r^2\frac{d\phi}{d\lambda},
\end{equation}
Substituting Eqs. (\ref{E}, \ref{L}) into Eq. (\ref{light-like}), we obtain the geodesic equation
\begin{equation}\label{0geo}
\left( \frac{dr}{d\lambda} \right)^2 = L^2\left[\frac{1}{b^2} - V_{\text{eff}}(r)\right] ,
\end{equation}
where
\begin{equation}
\label{0Veff}
V_{\text{eff}}(r) \equiv \frac{f(r)}{r^2},
\end{equation}
is the effective potential, and $b = L/E$ is the photon's impact parameter. Photon trajectories are determined by their impact parameter. The photon sphere corresponds to unstable circular null orbits at a radius $r_{\text{ph}}$. It can be determined by solving $V'_{\text{eff}}(r_{\text{ph}}) = 0$. The critical impact parameter $b_c$ at this radius is
\begin{equation}
b_c = \frac{r_{\text{ph}}}{\sqrt{f(r_{\text{ph}})}}.
\end{equation}

\begin{figure*}[ht]
    \centering
    \begin{minipage}[t]{0.48\textwidth}
        \centering
        \includegraphics[width=\linewidth]{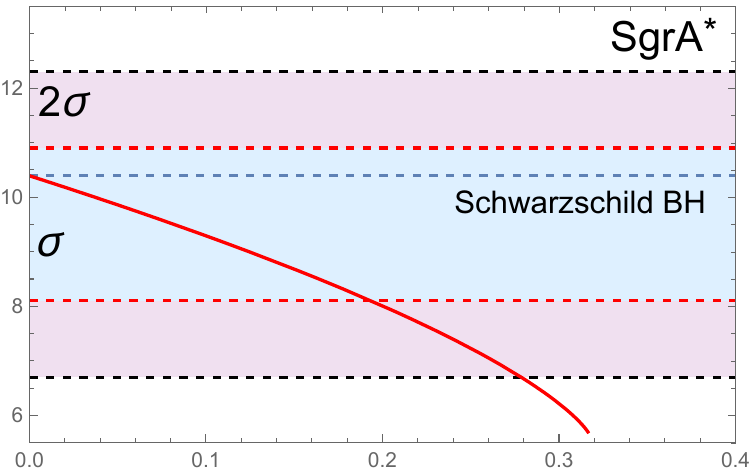}
    \end{minipage}
    \hfill
    \begin{minipage}[t]{0.48\textwidth}
        \centering
        \includegraphics[width=\linewidth]{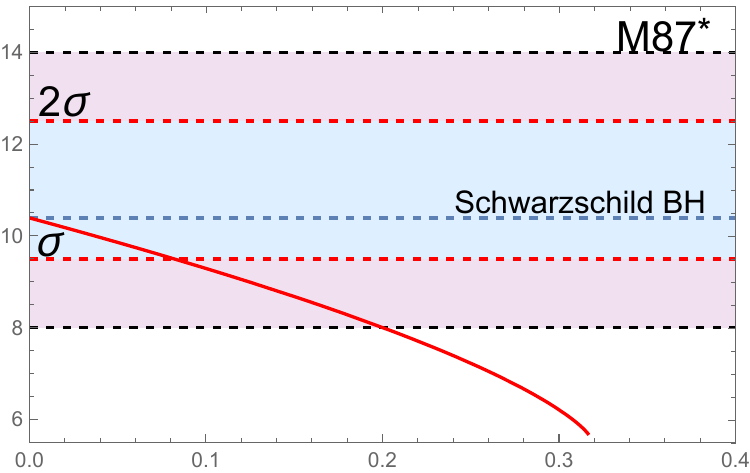}
    \end{minipage}
    \caption{The left panel shows the constraints on the model parameter from Sgr A* observational data. The right panel shows the constraints from M87* observational data. The shadow diameter for this black hole model is indicated by the solid red curve, and the Schwarzschild model is indicated by the dashed blue line. The blue and pink shaded regions represent the constraints on the shadow diameter from the EHT for Sgr A* and M87* at the $1\sigma$ and $2\sigma$ confidence levels, respectively. The gray region lies outside the $2\sigma$ range.}
    \label{SgrAM87}
\end{figure*}

Fig.~\ref{SgrAM87} displays the constraints on the metric parameter from EHT observations. For M87*, the boundary value for the parameter $a$ at the confidence level $1\sigma$ is $0.0819328$, and at the confidence level $2\sigma$ it is $0.20034$. For Sgr A*, the boundary value for the parameter $a$ at the confidence level $1\sigma$ is $0.1932$, and at the confidence level $2\sigma$ it is $0.279249$. These results indicate that the observational data for the two black holes impose different constraint ranges on the parameter $a$, with the constraints from Sgr A* being relatively looser.

\section{Motions of Massive Test Particles and Circular Orbits}
\label{MotionOfMassiveTestParticlesAndCircularOrbits}

\begin{figure}[htbp]
    \centering
    \includegraphics[width=\columnwidth]{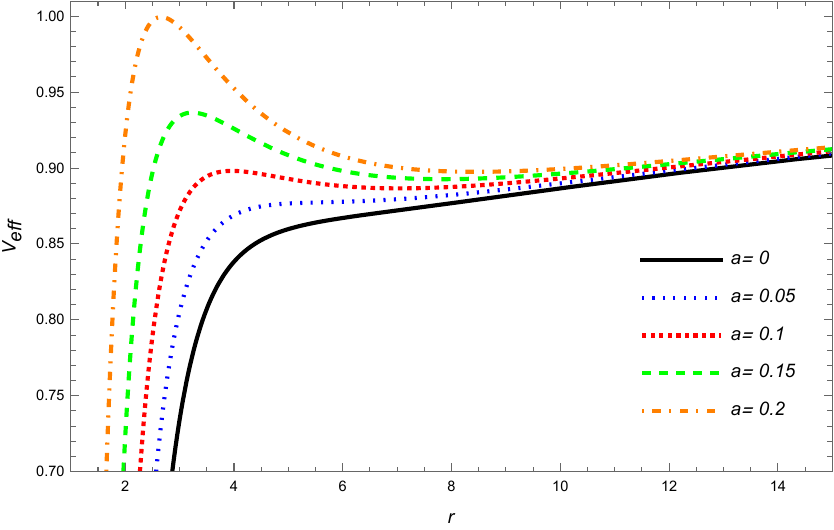}
    \caption{The $V_{\text{eff}}$ curves for different $a$, with $M=1$ and $L = 3.29$.}
    \label{Veff1}
\end{figure}

Through a similar discussion, we can also obtain the geodesic equation for a massive particle moving in the equatorial plane
\begin{equation}\label{geo}
\left( \frac{dr}{d\tau} \right)^2 = E^2 - V_{\text{eff}}(r; L),
\end{equation}
where $\tau$ is the proper time, and
\begin{equation}
\label{Veff}
V_{\text{eff}}(r; L) \equiv \left( 1 + \frac{L^2}{r^2} \right) f(r),
\end{equation}
is the effective potential. Fig.~\ref{Veff1} shows the variation of $V_{\text{eff}}(r; L)$ for different values of the parameter $a$. It can be seen that as $a$ increases, the maximum point of the curve tends to diverge, leading to a higher potential barrier.

Rewriting Eq. (\ref{geo}) using the chain rule $\frac{dr}{d\tau} = \frac{dr}{d\phi}\frac{d\phi}{d\tau}$, we obtain the equation for the particle's trajectory
\begin{equation}\label{r-phi}
\left( \frac{dr}{d\phi} \right)^2 = r^4 \left[ \frac{E^2}{L^2} - \frac{V_{\text{eff}}(r; L)}{L^2} \right].
\end{equation}
Circular orbits for particles are determined by the following conditions
\begin{equation}
V_{\text{eff}}(r; L) = E^2, \quad \frac{dV_{\text{eff}}(r; L)}{dr} = 0.
\end{equation}
From these conditions, we can obtain the energy $E$, the angular momentum $L$, and the angular velocity $\Omega$ for a particle in an equatorial circular orbit of radius $r$
\begin{equation}
\Omega = \frac{d\phi}{dt} = \sqrt{-\frac{g_{tt,r}}{g_{\phi\phi,r}}} = \sqrt{\frac{f'(r)}{2r}},
\end{equation}
\begin{equation}
E = -\frac{g_{tt}}{\sqrt{g_{tt} - g_{\phi\phi}\Omega^2}} = \sqrt{\frac{2f(r)^2}{2f(r) - r f'(r)}},
\end{equation}
\begin{equation}
L = \frac{g_{\phi\phi}\Omega}{\sqrt{g_{tt} - g_{\phi\phi}\Omega^2}} = \sqrt{\frac{r^3 f'(r)}{2f(r) - r f'(r)}}.
\end{equation}

In particular, the ISCO is the orbit satisfying the following conditions
\begin{equation}\label{Parameters of the ISCO}
V_{\text{eff}}(r; L) = E_{\text{ISCO}}^2, \quad \frac{dV_{\text{eff}}(r; L)}{dr} = 0, \quad \frac{d^2V_{\text{eff}}(r; L)}{dr^2} = 0.
\end{equation}
Using Eq. (\ref{Parameters of the ISCO}), the orbital parameters of the ISCO around the black hole can be found as
\begin{equation}\label{rISCO}
    r_{\text{ISCO}} = \frac{3f(r_{\text{ISCO}}) f'(r_{\text{ISCO}})}{2(f'(r_{\text{ISCO}}))^2 - f(r_{\text{ISCO}}) f''(r_{\text{ISCO}})},
\end{equation}
\begin{equation}\label{LISCO}
    L_{\text{ISCO}} = r_{\text{ISCO}}^{3/2} \sqrt{\frac{f'(r_{\text{ISCO}})}{2f(r_{\text{ISCO}}) - r_{\text{ISCO}} f'(r_{\text{ISCO}})}},
\end{equation}
\begin{equation}\label{EISCO}
    E_{\text{ISCO}} = \frac{f(r_{\text{ISCO}})}{\sqrt{f(r_{\text{ISCO}}) - \frac{1}{2}r_{\text{ISCO}} f'(r_{\text{ISCO}})}}.
\end{equation}

\begin{figure*}[htbp]
    \centering
    \begin{minipage}{0.32\textwidth}
        \centering
        \includegraphics[width=\linewidth]{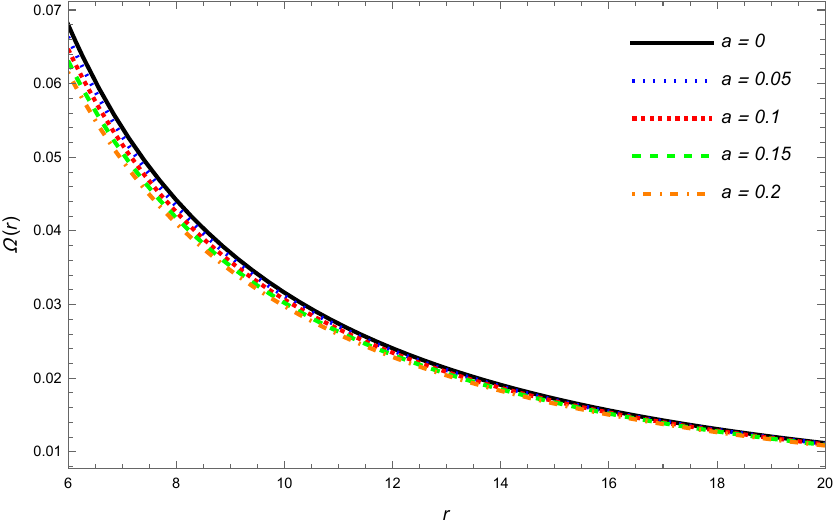}
    \end{minipage}
    \hfill
    \begin{minipage}{0.32\textwidth}
        \centering
        \includegraphics[width=\linewidth]{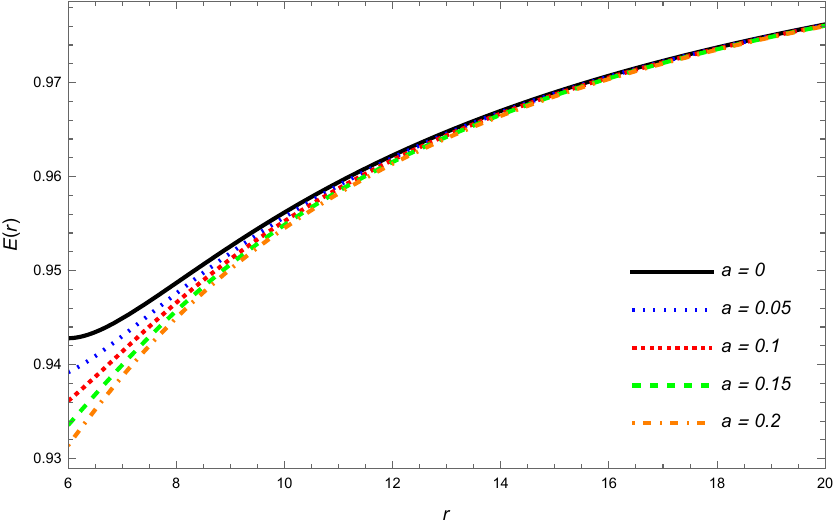}
    \end{minipage}
    \hfill
    \begin{minipage}{0.32\textwidth}
        \centering
        \includegraphics[width=\linewidth]{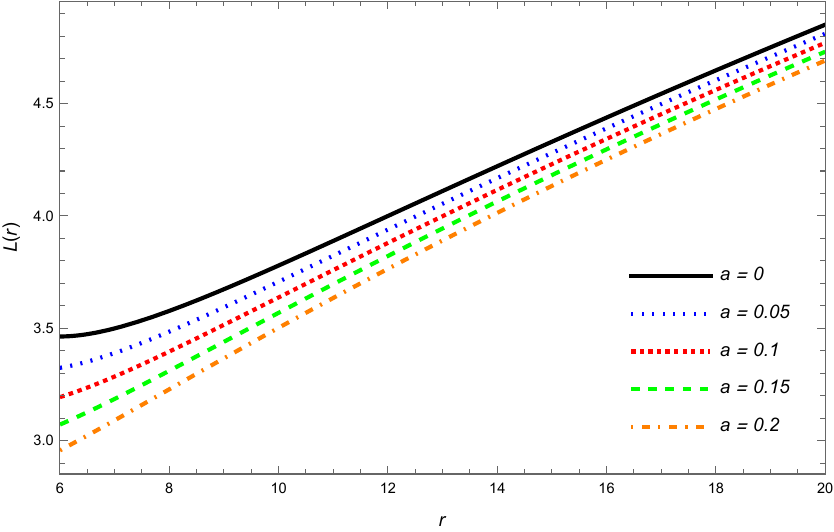}
    \end{minipage}
    \caption{Variation of circular orbit parameters, $\Omega$, $E$, $L$,  with $r$ for different values of the parameter $a$, with $M=1$ fixed.}
    \label{E,L,omega}
\end{figure*}

\begin{table}[h!]
    \centering
    \caption{Numerical results for the ISCO orbital parameters as functions of the metric parameter $a$, with $M=1$ fixed.}
    \label{ISCO}
    \begin{tabular}{l S[table-format=1.6] S[table-format=1.6] S[table-format=1.6] S[table-format=1.6]}
        \toprule
        $a$ & {$r_{\text{ISCO}}$} & {$\Omega_{\text{ISCO}}$} & {$E_{\text{ISCO}}$} & {$L_{\text{ISCO}}$} \\
        \midrule
        0 & 6. & 0.0680414 & 0.942809 & 3.4641 \\
        0.05 & 5.54138 & 0.0746133 & 0.938508 & 3.31485 \\
        0.1 & 5.06204 & 0.0827497 & 0.933276 & 3.15404 \\
        0.15 & 4.55439 & 0.0931991 & 0.926689 & 2.9781 \\
        0.2 & 4.00497 & 0.10738 & 0.917965 & 2.78092 \\
        \bottomrule
    \end{tabular}
\end{table}

Fig.~\ref{E,L,omega} shows the variation curves of circular orbit parameters for different values of the metric parameter $a$. It can be seen that parameter's changes have a significant impact on the orbital parameters in the strong-field region, but both the angular velocity and the energy tend towards the Schwarzschild black hole case in the weak-field region. Meanwhile, for larger values of $a$, circular orbits at the same radius have a smaller angular momentum.

Table~\ref{ISCO} presents the calculated results for ISCO-related parameters as the metric parameter $a$ varies. The analysis shows that as the parameter $a$ increases, the ISCO exhibits systematic trends: the orbital radius $r_{\text{ISCO}}$ noticeably contracts towards the black hole horizon; the corresponding orbital angular velocity $\Omega_{\text{ISCO}}$ increases significantly. At the same time, the particle's energy $E_{\text{ISCO}}$ on this orbit decreases slightly, while the angular momentum $L_{\text{ISCO}}$ shows a more rapid decreasing trend. These variation patterns reflect the direct influence of the metric parameter $a$ on the spacetime geometry and the structure of the gravitational field around the black hole, thus modulating the existence range and dynamical characteristics of stable orbits.

\section{Observed Flux, Redshift, and Black Hole Optical Appearance}
\label{ObservedFluxRedshiftAndBlackHoleOpticalAppearance}
Based on the relativistic thin accretion disk model proposed in \cite{Page:1974he, Collodel:2021gxu}, the local radiation flux on the disk surface can be expressed as
\begin{equation}
\label{F(r)}
F(r) = -\frac{\dot{M} \Omega_{,r}}{4\pi\sqrt{-g/g_{\theta\theta}}(E - \Omega L)^2}\int_{r_{\text{ISCO}}}^{r} (E - \Omega L)L_{,r}\,dr,
\end{equation}
where $g$ is the metric determinant, $\dot{M}$ is the mass accretion rate. This model is applicable to geometrically thin, optically thick accretion disks, with the integration extending from the ISCO to the radius $r$, representing the cumulative contribution of all radiation emitted from the inner edge to this radius.

Due to the gravitational redshift effect in strong gravitational fields and the Doppler effect caused by the orbital motion of the accretion disk material, the radiation flux received by a distant observer $F_{\text{obs}}$ differs from the locally emitted flux $F(r)$, they are related through the total redshift factor $z$
\begin{equation}
\label{Fobs}
F_{\text{obs}} = \frac{F(r)}{(1 + z)^4}.
\end{equation}
The total redshift factor $z$ combines the contributions of gravitational redshift and Doppler redshift and is given by \cite{Luminet:1979nyg, Huang:2023ilm}
\begin{equation}
\label{z}
1 + z = \frac{1 + \Omega b\sin\theta_0\cos\alpha}{\sqrt{-g_{tt} - g_{\phi\phi}\Omega^2}},
\end{equation}
where $\theta_0$ is the observer's inclination angle, and $\alpha$ is the polar angle of the photon's position on the observer's image plane. The denominator $\sqrt{-g_{tt} - g_{\phi\phi}\Omega^2}$ originates from gravitational redshift, while the numerator $1 + \Omega b\sin\theta_0\cos\alpha$ embodies the coupling between the Doppler effect and the observational geometry.

Combining equations \eqref{F(r)}, \eqref{Fobs} and \eqref{z}, we obtain the radiation flux distribution received by the observer
\begin{equation}
F_{\text{obs}} = \frac{\displaystyle -\frac{\dot{M}\Omega_{,r}}{4\pi\sqrt{-g/g_{\theta\theta}}(E - \Omega L)^2}\int_{r_{\text{ISCO}}}^{r}(E - \Omega L)L_{,r}\,dr}{\displaystyle \left(\frac{1 + \Omega b\sin\theta_0\cos\alpha}{\sqrt{-g_{tt} - g_{\phi\phi}\Omega^2}}\right)^4}.
\end{equation}

\begin{figure*}[htbp]
    \centering
    \newsavebox{\mybox} % For measuring main figure height

    % ------------------ First row ------------------
    \begin{subfigure}[b]{0.24\textwidth}
        \centering
        \sbox{\mybox}{\includegraphics[width=0.7\linewidth]{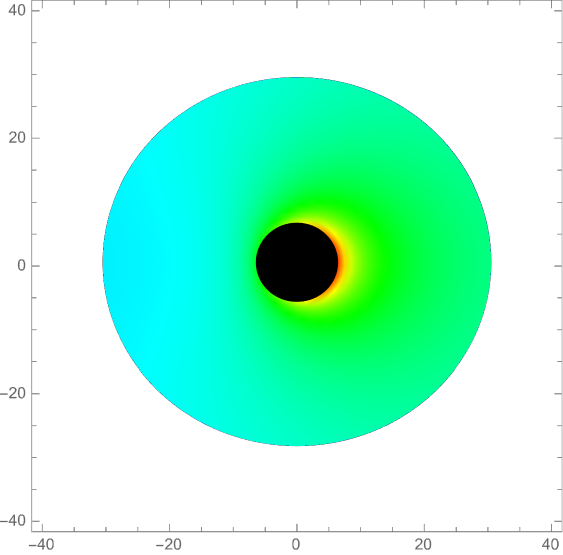}}%
        \usebox{\mybox}%
        \hfill
        \includegraphics[height=\ht\mybox]{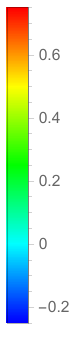}%
        \caption{\scriptsize $a=0.05,\theta_0=20^\circ$}
    \end{subfigure}
    \hfill
    \begin{subfigure}[b]{0.24\textwidth}
        \centering
        \sbox{\mybox}{\includegraphics[width=0.7\linewidth]{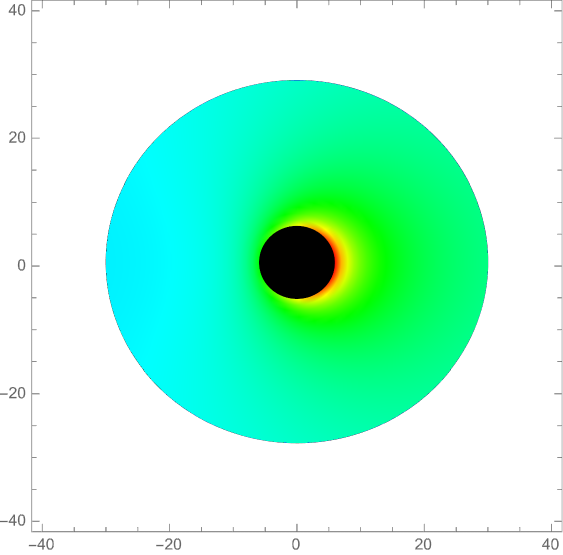}}%
        \usebox{\mybox}%
        \hfill
        \includegraphics[height=\ht\mybox]{color/redshift1.pdf}%
        \caption{\scriptsize $a=0.1,\theta_0=20^\circ$}
    \end{subfigure}
    \hfill
    \begin{subfigure}[b]{0.24\textwidth}
        \centering
        \sbox{\mybox}{\includegraphics[width=0.7\linewidth]{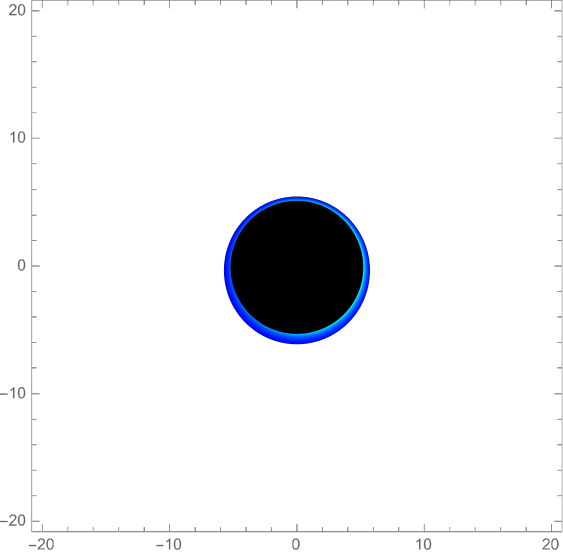}}%
        \usebox{\mybox}%
        \hfill
        \includegraphics[height=\ht\mybox]{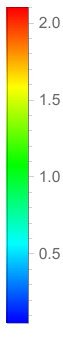}%
        \caption{\scriptsize $a=0.05,\theta_0=20^\circ$}
    \end{subfigure}
    \hfill
    \begin{subfigure}[b]{0.24\textwidth}
        \centering
        \sbox{\mybox}{\includegraphics[width=0.7\linewidth]{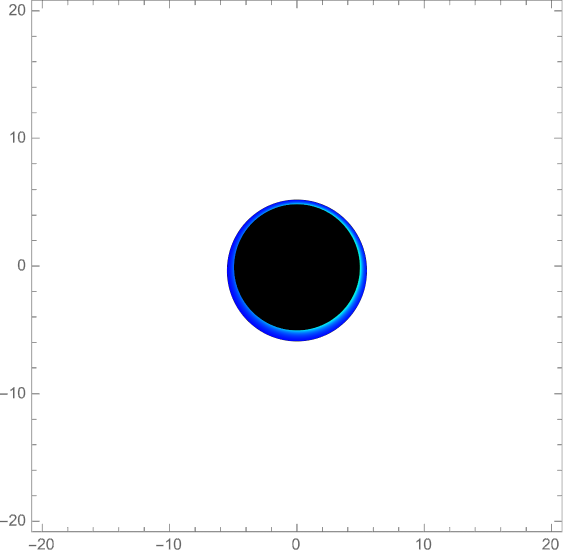}}%
        \usebox{\mybox}%
        \hfill
        \includegraphics[height=\ht\mybox]{color/redshift2.pdf}%
        \caption{\scriptsize $a=0.1,\theta_0=20^\circ$}
    \end{subfigure}

    \vspace{0.5cm}

    % ------------------ Second row ------------------
    \begin{subfigure}[b]{0.24\textwidth}
        \centering
        \sbox{\mybox}{\includegraphics[width=0.7\linewidth]{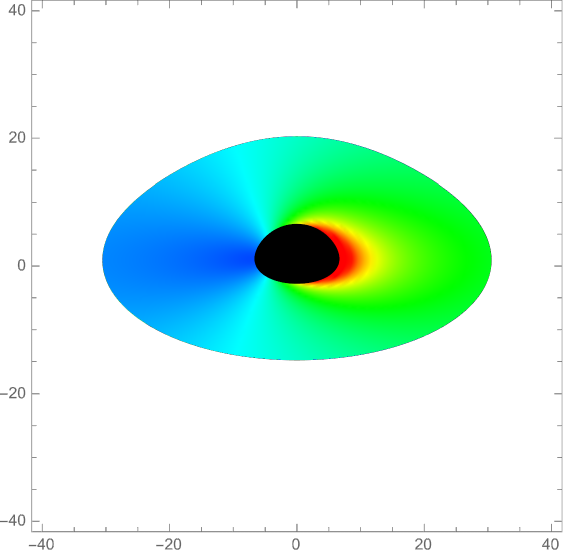}}%
        \usebox{\mybox}%
        \hfill
        \includegraphics[height=\ht\mybox]{color/redshift1.pdf}%
        \caption{\scriptsize $a=0.05,\theta_0=60^\circ$}
    \end{subfigure}
    \hfill
    \begin{subfigure}[b]{0.24\textwidth}
        \centering
        \sbox{\mybox}{\includegraphics[width=0.7\linewidth]{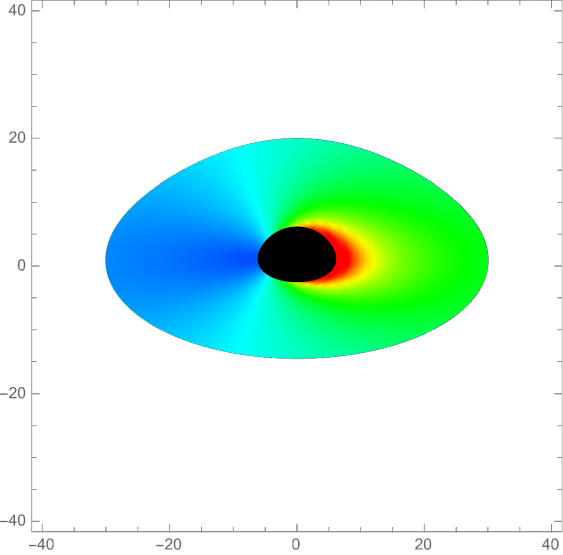}}%
        \usebox{\mybox}%
        \hfill
        \includegraphics[height=\ht\mybox]{color/redshift1.pdf}%
        \caption{\scriptsize $a=0.1,\theta_0=60^\circ$}
    \end{subfigure}
    \hfill
    \begin{subfigure}[b]{0.24\textwidth}
        \centering
        \sbox{\mybox}{\includegraphics[width=0.7\linewidth]{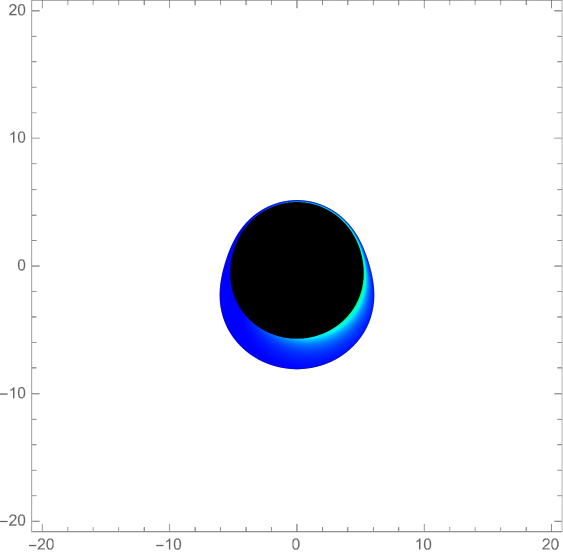}}%
        \usebox{\mybox}%
        \hfill
        \includegraphics[height=\ht\mybox]{color/redshift2.pdf}%
        \caption{\scriptsize $a=0.05,\theta_0=60^\circ$}
    \end{subfigure}
    \hfill
    \begin{subfigure}[b]{0.24\textwidth}
        \centering
        \sbox{\mybox}{\includegraphics[width=0.7\linewidth]{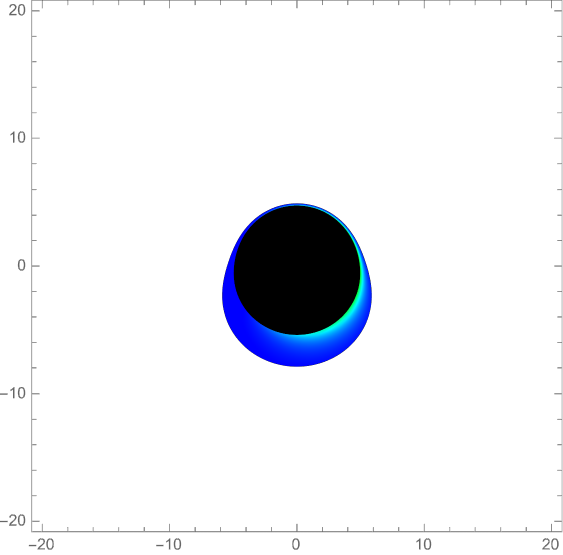}}%
        \usebox{\mybox}%
        \hfill
        \includegraphics[height=\ht\mybox]{color/redshift2.pdf}%
        \caption{\scriptsize $a=0.1,\theta_0=60^\circ$}
    \end{subfigure}

    \vspace{0.5cm}

    % ------------------ Third row ------------------
    \begin{subfigure}[b]{0.24\textwidth}
        \centering
        \sbox{\mybox}{\includegraphics[width=0.7\linewidth]{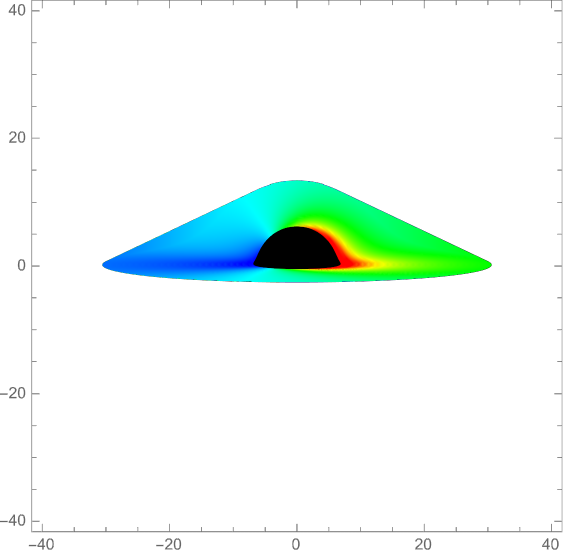}}%
        \usebox{\mybox}%
        \hfill
        \includegraphics[height=\ht\mybox]{color/redshift1.pdf}%
        \caption{\scriptsize $a=0.05,\theta_0=85^\circ$}
    \end{subfigure}
    \hfill
    \begin{subfigure}[b]{0.24\textwidth}
        \centering
        \sbox{\mybox}{\includegraphics[width=0.7\linewidth]{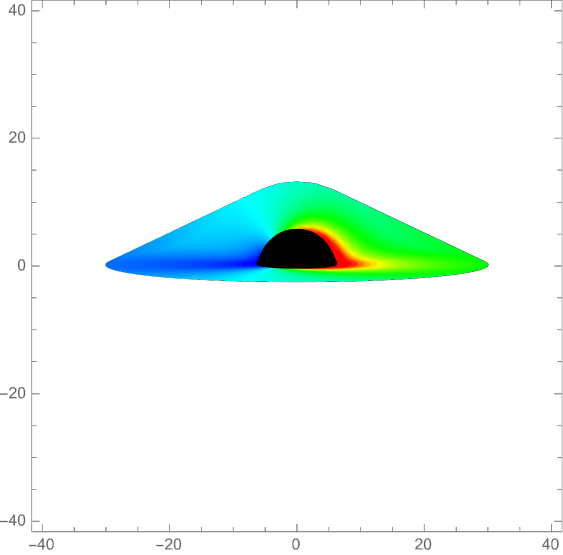}}%
        \usebox{\mybox}%
        \hfill
        \includegraphics[height=\ht\mybox]{color/redshift1.pdf}%
        \caption{\scriptsize $a=0.1,\theta_0=85^\circ$}
    \end{subfigure}
    \hfill
    \begin{subfigure}[b]{0.24\textwidth}
        \centering
        \sbox{\mybox}{\includegraphics[width=0.7\linewidth]{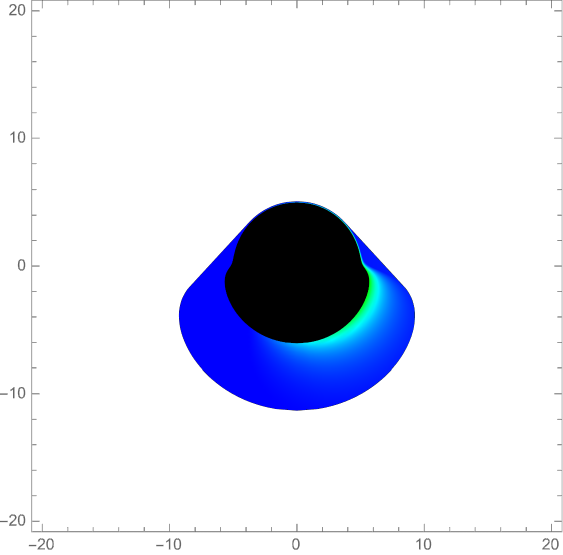}}%
        \usebox{\mybox}%
        \hfill
        \includegraphics[height=\ht\mybox]{color/redshift2.pdf}%
        \caption{\scriptsize $a=0.05,\theta_0=85^\circ$}
    \end{subfigure}
    \hfill
    \begin{subfigure}[b]{0.24\textwidth}
        \centering
        \sbox{\mybox}{\includegraphics[width=0.7\linewidth]{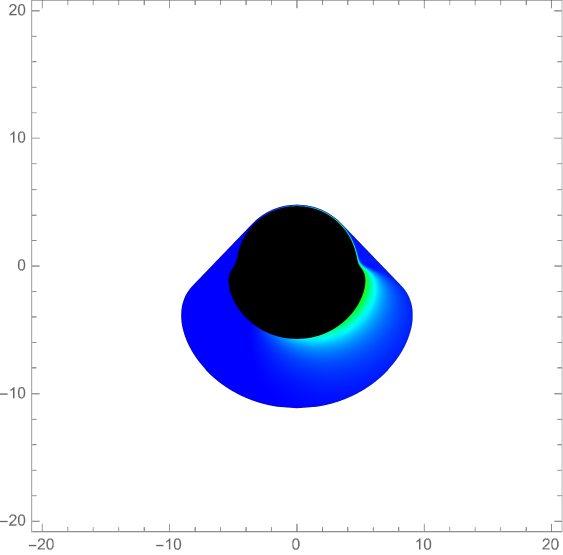}}%
        \usebox{\mybox}%
        \hfill
        \includegraphics[height=\ht\mybox]{color/redshift2.pdf}%
        \caption{\scriptsize $a=0.1,\theta_0=85^\circ$}
    \end{subfigure}

    \caption{The influence of the parameter $a$ on the direct and secondary images of the redshift factor $z$ for the black hole accretion disk at different viewing angles. The outer edge of the accretion disk is at $r=30M$, with $M=1$ fixed.}
    \label{Z1}
\end{figure*}

\begin{figure*}[htbp]
    \centering

    % ------------------ First row (θ₀=20°) ------------------
    \begin{subfigure}[b]{0.24\textwidth}
        \centering
        \sbox{\mybox}{\includegraphics[width=0.7\linewidth]{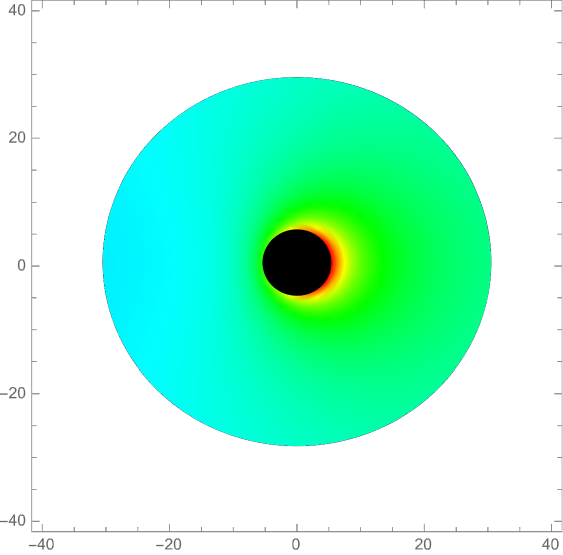}}%
        \usebox{\mybox}%
        \hfill
        \includegraphics[height=\ht\mybox]{color/redshift1.pdf}%
        \caption{\scriptsize $a=0.15,\theta_0=20^\circ$}
    \end{subfigure}
    \hfill
    \begin{subfigure}[b]{0.24\textwidth}
        \centering
        \sbox{\mybox}{\includegraphics[width=0.7\linewidth]{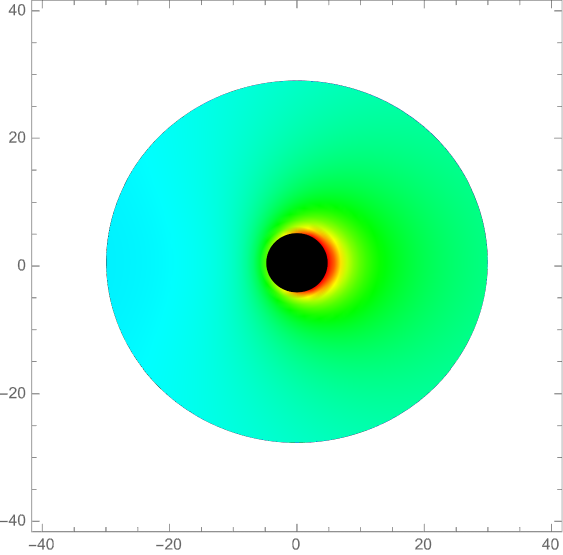}}%
        \usebox{\mybox}%
        \hfill
        \includegraphics[height=\ht\mybox]{color/redshift1.pdf}%
        \caption{\scriptsize $a=0.2,\theta_0=20^\circ$}
    \end{subfigure}
    \hfill
    \begin{subfigure}[b]{0.24\textwidth}
        \centering
        \sbox{\mybox}{\includegraphics[width=0.7\linewidth]{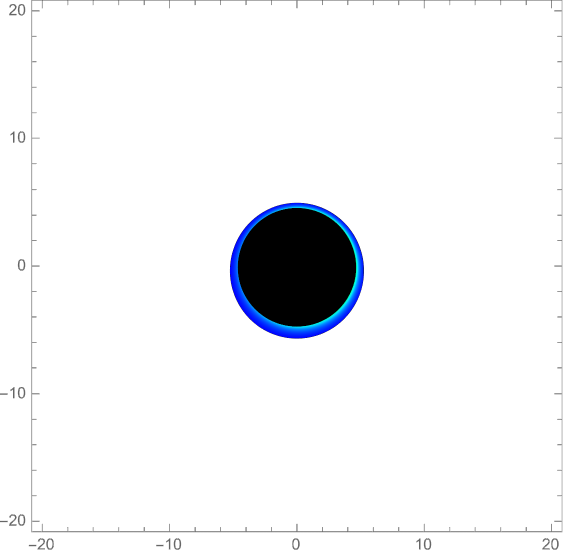}}%
        \usebox{\mybox}%
        \hfill
        \includegraphics[height=\ht\mybox]{color/redshift2.pdf}%
        \caption{\scriptsize $a=0.15,\theta_0=20^\circ$}
    \end{subfigure}
    \hfill
    \begin{subfigure}[b]{0.24\textwidth}
        \centering
        \sbox{\mybox}{\includegraphics[width=0.7\linewidth]{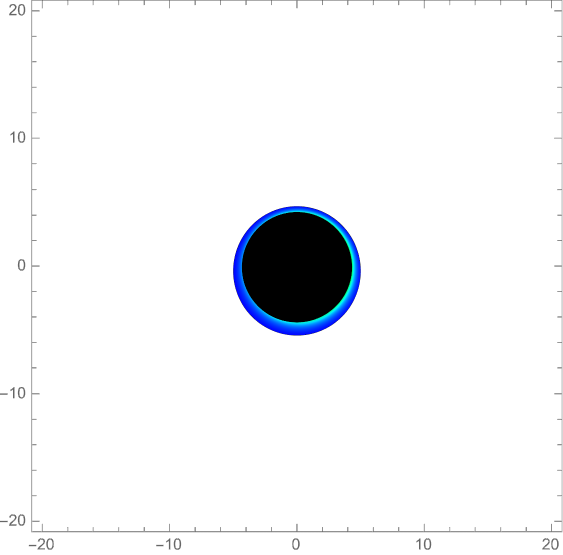}}%
        \usebox{\mybox}%
        \hfill
        \includegraphics[height=\ht\mybox]{color/redshift2.pdf}%
        \caption{\scriptsize $a=0.2,\theta_0=20^\circ$}
    \end{subfigure}

    \vspace{0.5cm}

    % ------------------ Second row (θ₀=60°) ------------------
    \begin{subfigure}[b]{0.24\textwidth}
        \centering
        \sbox{\mybox}{\includegraphics[width=0.7\linewidth]{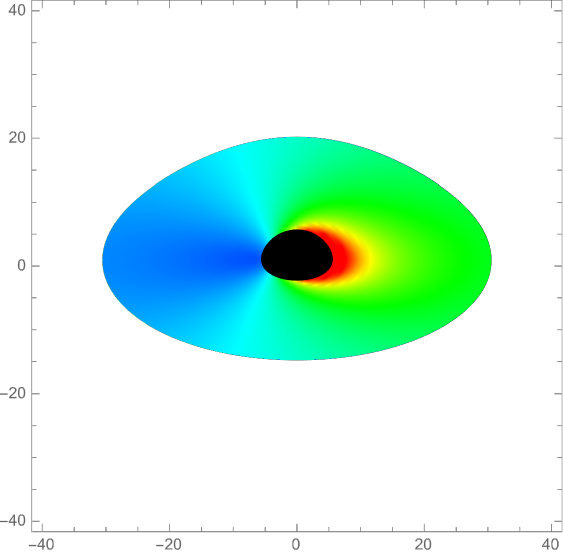}}%
        \usebox{\mybox}%
        \hfill
        \includegraphics[height=\ht\mybox]{color/redshift1.pdf}%
        \caption{\scriptsize $a=0.15,\theta_0=60^\circ$}
    \end{subfigure}
    \hfill
    \begin{subfigure}[b]{0.24\textwidth}
        \centering
        \sbox{\mybox}{\includegraphics[width=0.7\linewidth]{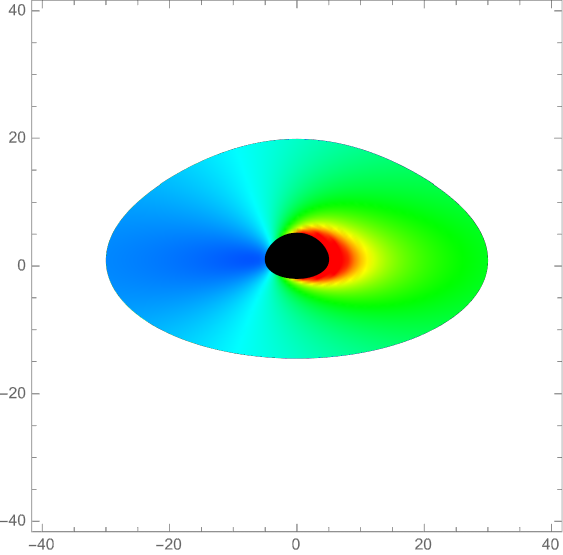}}%
        \usebox{\mybox}%
        \hfill
        \includegraphics[height=\ht\mybox]{color/redshift1.pdf}%
        \caption{\scriptsize $a=0.2,\theta_0=60^\circ$}
    \end{subfigure}
    \hfill
    \begin{subfigure}[b]{0.24\textwidth}
        \centering
        \sbox{\mybox}{\includegraphics[width=0.7\linewidth]{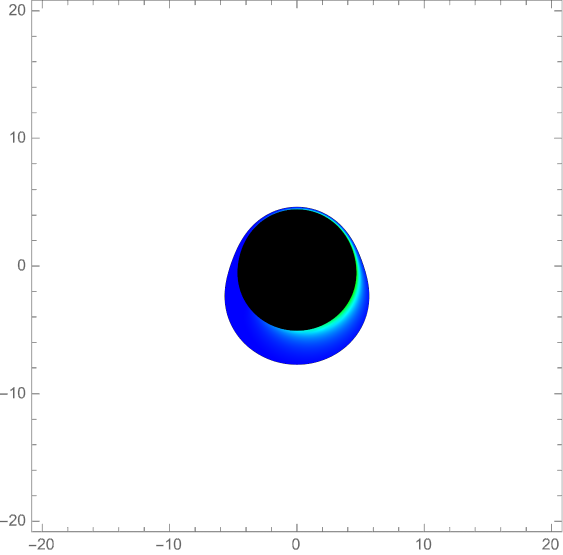}}%
        \usebox{\mybox}%
        \hfill
        \includegraphics[height=\ht\mybox]{color/redshift2.pdf}%
        \caption{\scriptsize $a=0.15,\theta_0=60^\circ$}
    \end{subfigure}
    \hfill
    \begin{subfigure}[b]{0.24\textwidth}
        \centering
        \sbox{\mybox}{\includegraphics[width=0.7\linewidth]{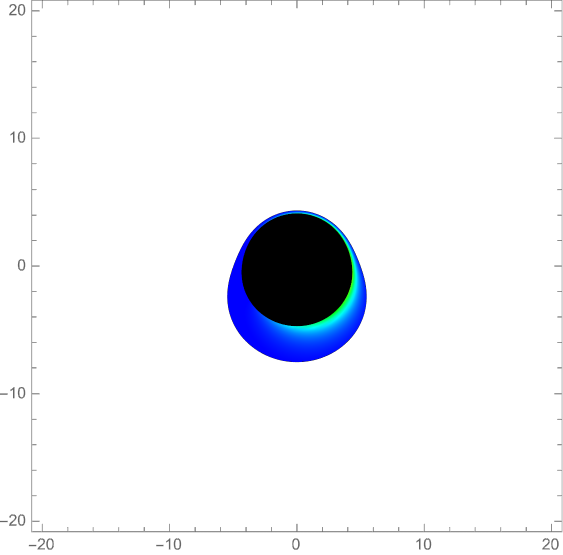}}%
        \usebox{\mybox}%
        \hfill
        \includegraphics[height=\ht\mybox]{color/redshift2.pdf}%
        \caption{\scriptsize $a=0.2,\theta_0=60^\circ$}
    \end{subfigure}

    \vspace{0.5cm}

    % ------------------ Third row (θ₀=85°) ------------------
    \begin{subfigure}[b]{0.24\textwidth}
        \centering
        \sbox{\mybox}{\includegraphics[width=0.7\linewidth]{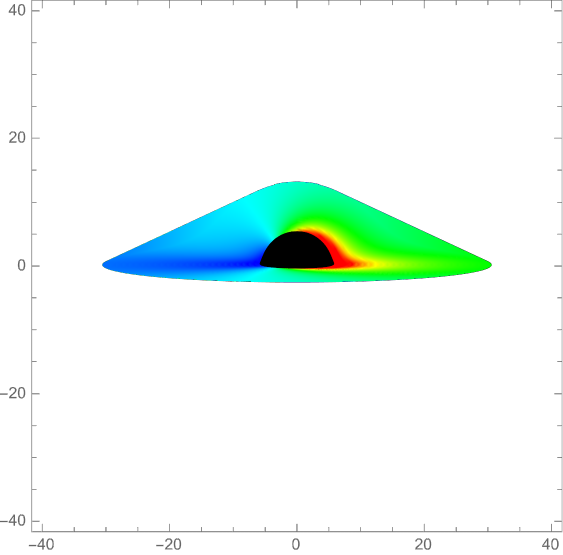}}%
        \usebox{\mybox}%
        \hfill
        \includegraphics[height=\ht\mybox]{color/redshift1.pdf}%
        \caption{\scriptsize $a=0.15,\theta_0=85^\circ$}
    \end{subfigure}
    \hfill
    \begin{subfigure}[b]{0.24\textwidth}
        \centering
        \sbox{\mybox}{\includegraphics[width=0.7\linewidth]{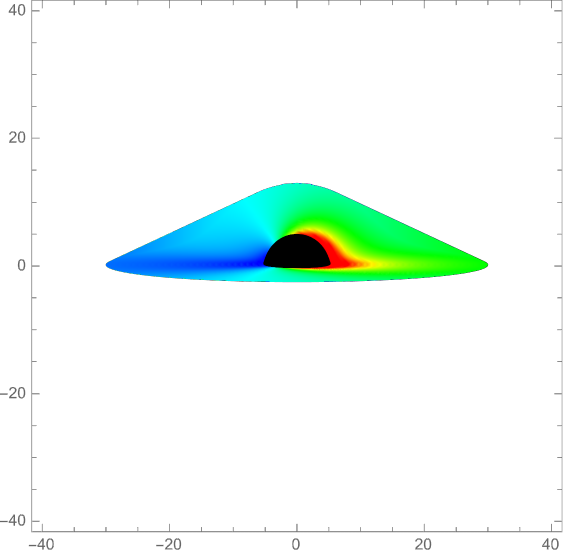}}%
        \usebox{\mybox}%
        \hfill
        \includegraphics[height=\ht\mybox]{color/redshift1.pdf}%
        \caption{\scriptsize $a=0.2,\theta_0=85^\circ$}
    \end{subfigure}
    \hfill
    \begin{subfigure}[b]{0.24\textwidth}
        \centering
        \sbox{\mybox}{\includegraphics[width=0.7\linewidth]{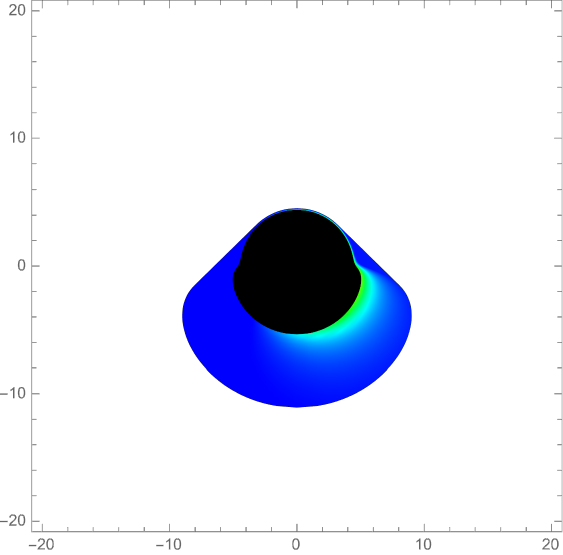}}%
        \usebox{\mybox}%
        \hfill
        \includegraphics[height=\ht\mybox]{color/redshift2.pdf}%
        \caption{\scriptsize $a=0.15,\theta_0=85^\circ$}
    \end{subfigure}
    \hfill
    \begin{subfigure}[b]{0.24\textwidth}
        \centering
        \sbox{\mybox}{\includegraphics[width=0.7\linewidth]{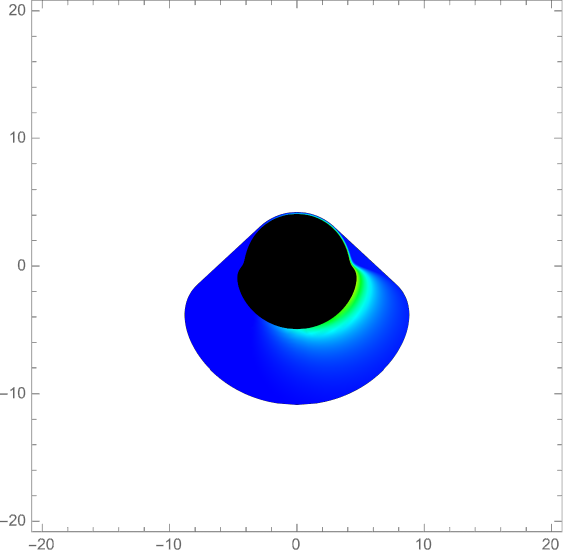}}%
        \usebox{\mybox}%
        \hfill
        \includegraphics[height=\ht\mybox]{color/redshift2.pdf}%
        \caption{\scriptsize $a=0.2,\theta_0=85^\circ$}
    \end{subfigure}

    \caption{The influence of the parameter $a$ on the direct and secondary images of the redshift factor $z$ for the black hole accretion disk at different viewing angles. The outer edge of the accretion disk is at $r=30M$, with $M=1$ fixed.}
    \label{Z2}
\end{figure*}

Figs.~\ref{Z1} and~\ref{Z2} illustrate the influence of the parameter $a$ on the redshift factor $z$ for the direct and secondary images of the black hole accretion disk at different viewing angles.
When the viewing angle $\theta_0 = 0$, the spacetime exhibits spherical symmetry, the Doppler redshift effect vanishes, and the gravitational redshift of the disk structure shows a rotationally symmetric distribution. As the viewing angle increases, the Doppler redshift effect becomes significantly enhanced: on one side of the disk (the part where material moves towards the observer), the Doppler blueshift partially cancels the gravitational redshift, causing the total redshift factor $z$ to decrease; on the other side, the Doppler redshift adds to the gravitational redshift, causing $z$ to increase. This asymmetry intensifies markedly with increasing $\theta_0$, resulting in a pronounced left-right asymmetry in the redshift factor image.

\begin{figure*}[htbp]
    \centering

    % First row (θ₀=20°)
    \begin{subfigure}[b]{0.24\textwidth}
        \centering
        \sbox{\mybox}{\includegraphics[width=0.7\linewidth]{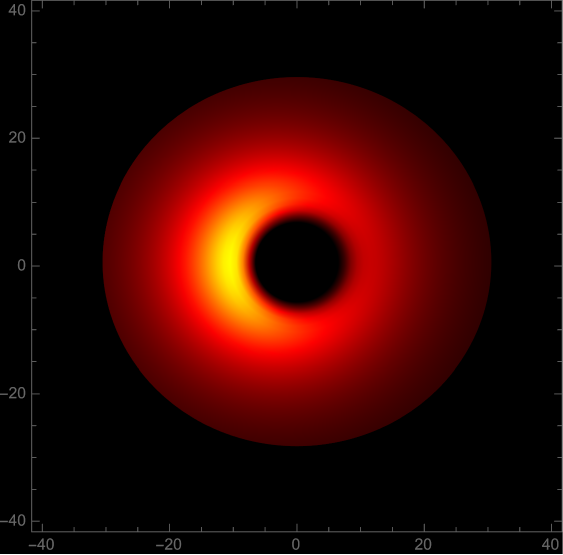}}%
        \usebox{\mybox}%
        \hfill
        \includegraphics[height=\ht\mybox]{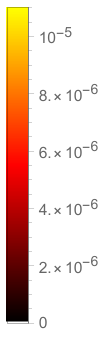}%
        \caption{\scriptsize $a=0.05,\theta_0=20^\circ$}
    \end{subfigure}
    \hfill
    \begin{subfigure}[b]{0.24\textwidth}
        \centering
        \sbox{\mybox}{\includegraphics[width=0.7\linewidth]{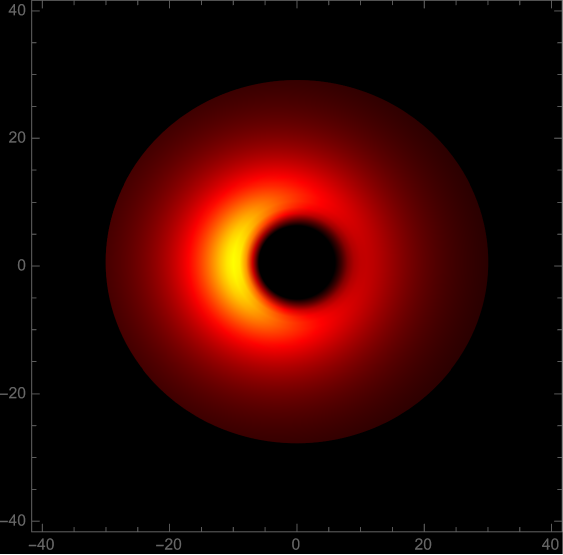}}%
        \usebox{\mybox}%
        \hfill
        \includegraphics[height=\ht\mybox]{color/acc-disk1.pdf}%
        \caption{\scriptsize $a=0.1,\theta_0=20^\circ$}
    \end{subfigure}
    \hfill
    \begin{subfigure}[b]{0.24\textwidth}
        \centering
        \sbox{\mybox}{\includegraphics[width=0.7\linewidth]{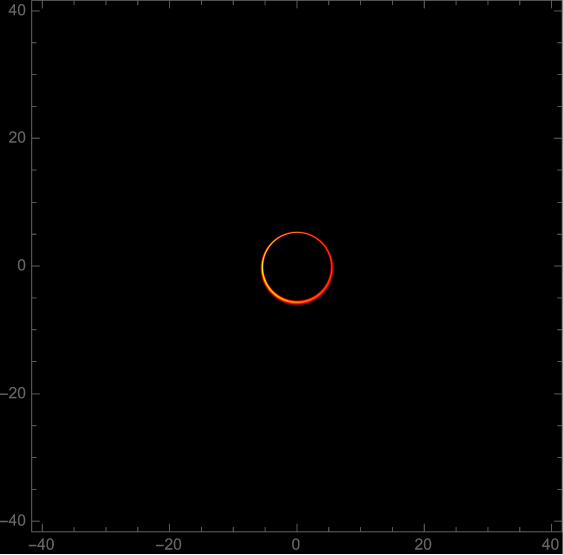}}%
        \usebox{\mybox}%
        \hfill
        \includegraphics[height=\ht\mybox]{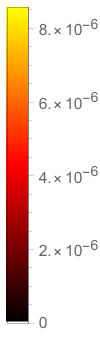}%
        \caption{\scriptsize $a=0.05,\theta_0=20^\circ$}
    \end{subfigure}
    \hfill
    \begin{subfigure}[b]{0.24\textwidth}
        \centering
        \sbox{\mybox}{\includegraphics[width=0.7\linewidth]{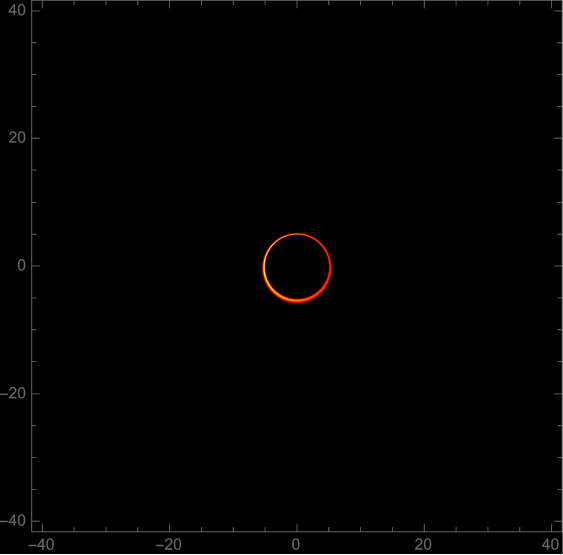}}%
        \usebox{\mybox}%
        \hfill
        \includegraphics[height=\ht\mybox]{color/acc-disk2.pdf}%
        \caption{\scriptsize $a=0.1,\theta_0=20^\circ$}
    \end{subfigure}

    \vspace{0.5cm}

    % Second row (θ₀=60°)
    \begin{subfigure}[b]{0.24\textwidth}
        \centering
        \sbox{\mybox}{\includegraphics[width=0.7\linewidth]{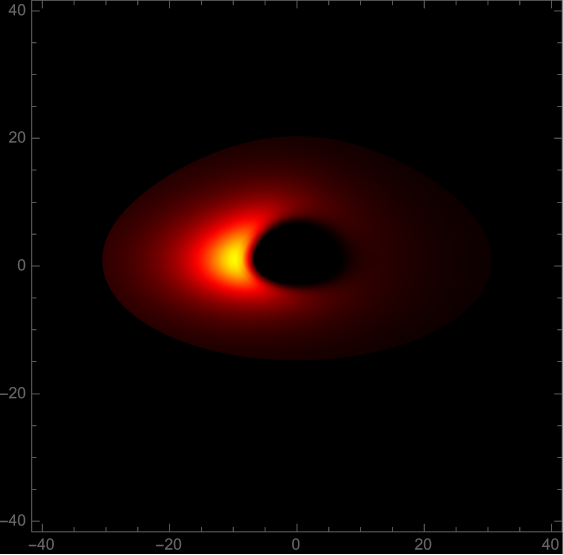}}%
        \usebox{\mybox}%
        \hfill
        \includegraphics[height=\ht\mybox]{color/acc-disk1.pdf}%
        \caption{\scriptsize $a=0.05,\theta_0=60^\circ$}
    \end{subfigure}
    \hfill
    \begin{subfigure}[b]{0.24\textwidth}
        \centering
        \sbox{\mybox}{\includegraphics[width=0.7\linewidth]{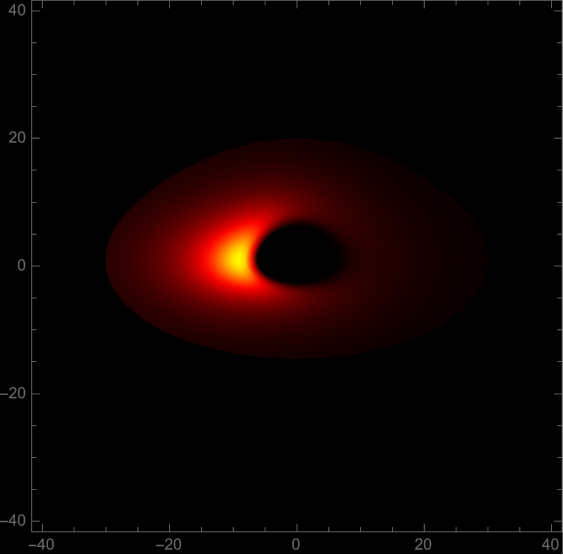}}%
        \usebox{\mybox}%
        \hfill
        \includegraphics[height=\ht\mybox]{color/acc-disk1.pdf}%
        \caption{\scriptsize $a=0.1,\theta_0=60^\circ$}
    \end{subfigure}
    \hfill
    \begin{subfigure}[b]{0.24\textwidth}
        \centering
        \sbox{\mybox}{\includegraphics[width=0.7\linewidth]{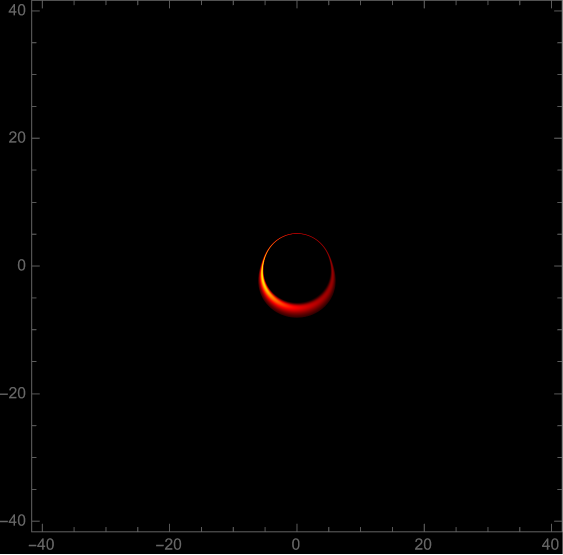}}%
        \usebox{\mybox}%
        \hfill
        \includegraphics[height=\ht\mybox]{color/acc-disk2.pdf}%
        \caption{\scriptsize $a=0.05,\theta_0=60^\circ$}
    \end{subfigure}
    \hfill
    \begin{subfigure}[b]{0.24\textwidth}
        \centering
        \sbox{\mybox}{\includegraphics[width=0.7\linewidth]{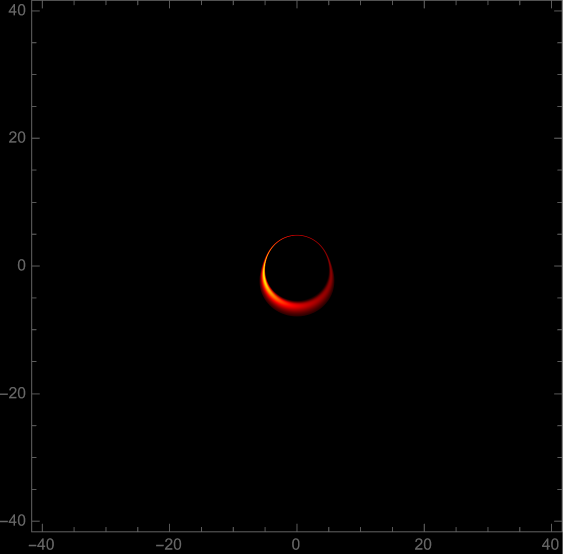}}%
        \usebox{\mybox}%
        \hfill
        \includegraphics[height=\ht\mybox]{color/acc-disk2.pdf}%
        \caption{\scriptsize $a=0.1,\theta_0=60^\circ$}
    \end{subfigure}

    \vspace{0.5cm}

    % Third row (θ₀=85°)
    \begin{subfigure}[b]{0.24\textwidth}
        \centering
        \sbox{\mybox}{\includegraphics[width=0.7\linewidth]{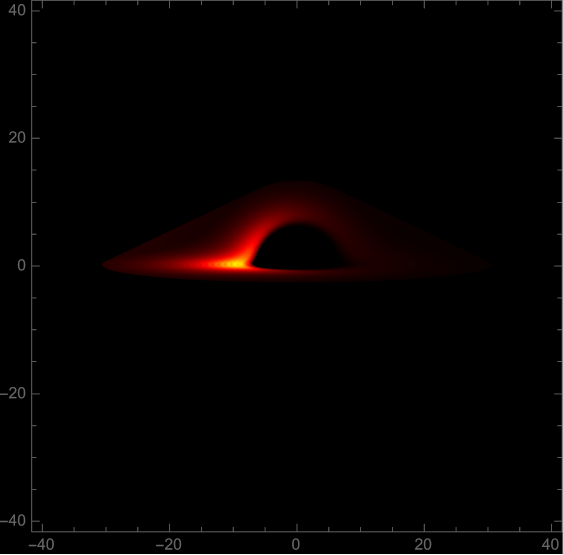}}%
        \usebox{\mybox}%
        \hfill
        \includegraphics[height=\ht\mybox]{color/acc-disk1.pdf}%
        \caption{\scriptsize $a=0.05,\theta_0=85^\circ$}
    \end{subfigure}
    \hfill
    \begin{subfigure}[b]{0.24\textwidth}
        \centering
        \sbox{\mybox}{\includegraphics[width=0.7\linewidth]{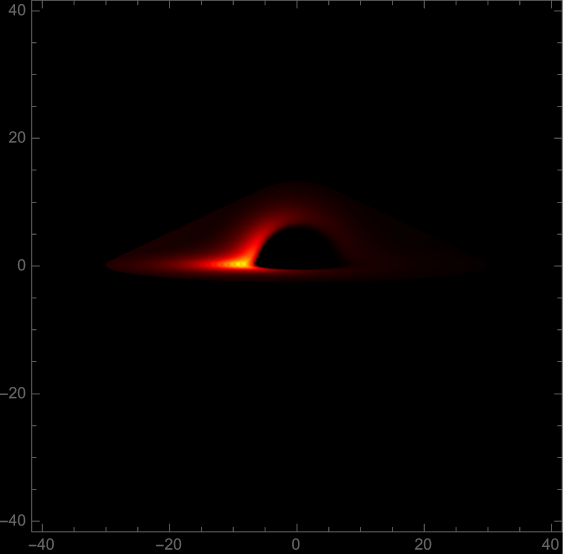}}%
        \usebox{\mybox}%
        \hfill
        \includegraphics[height=\ht\mybox]{color/acc-disk1.pdf}%
        \caption{\scriptsize $a=0.1,\theta_0=85^\circ$}
    \end{subfigure}
    \hfill
    \begin{subfigure}[b]{0.24\textwidth}
        \centering
        \sbox{\mybox}{\includegraphics[width=0.7\linewidth]{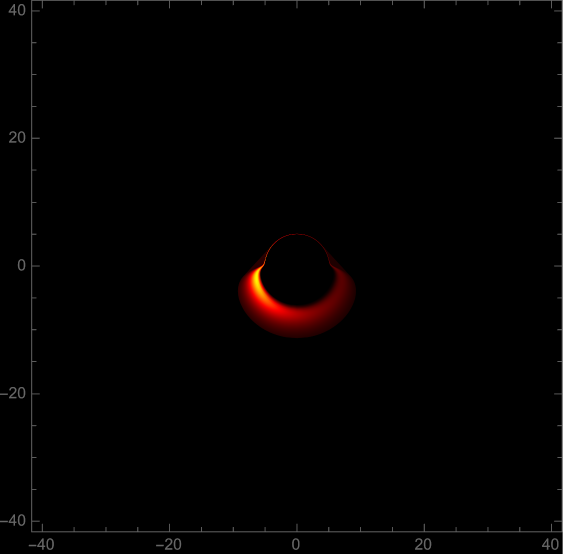}}%
        \usebox{\mybox}%
        \hfill
        \includegraphics[height=\ht\mybox]{color/acc-disk2.pdf}%
        \caption{\scriptsize $a=0.05,\theta_0=85^\circ$}
    \end{subfigure}
    \hfill
    \begin{subfigure}[b]{0.24\textwidth}
        \centering
        \sbox{\mybox}{\includegraphics[width=0.7\linewidth]{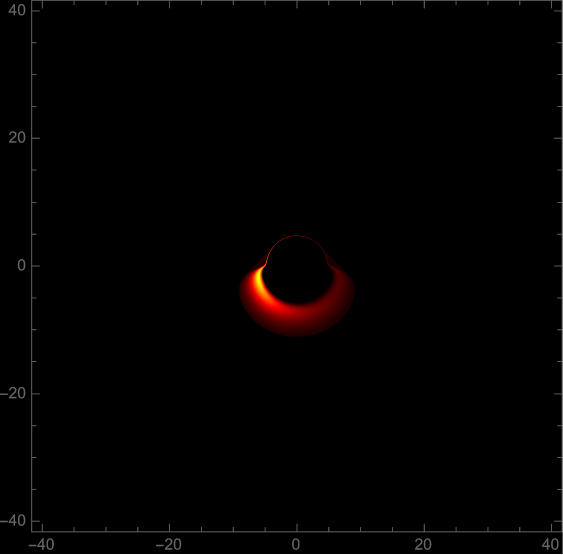}}%
        \usebox{\mybox}%
        \hfill
        \includegraphics[height=\ht\mybox]{color/acc-disk2.pdf}%
        \caption{\scriptsize $a=0.1,\theta_0=85^\circ$}
    \end{subfigure}

    \caption{The influence of the parameter $a$ on the direct and secondary images of the observed radiative flux $F_{\text{obs}}$ at different viewing angles. The outer edge of the accretion disk is at $r=30M$, with $M=1$ fixed.}
    \label{F1}
\end{figure*}

\begin{figure*}[htbp]
    \centering  

    % First row (θ₀=20°)
    \begin{subfigure}[b]{0.24\textwidth}
        \centering
        \sbox{\mybox}{\includegraphics[width=0.7\linewidth]{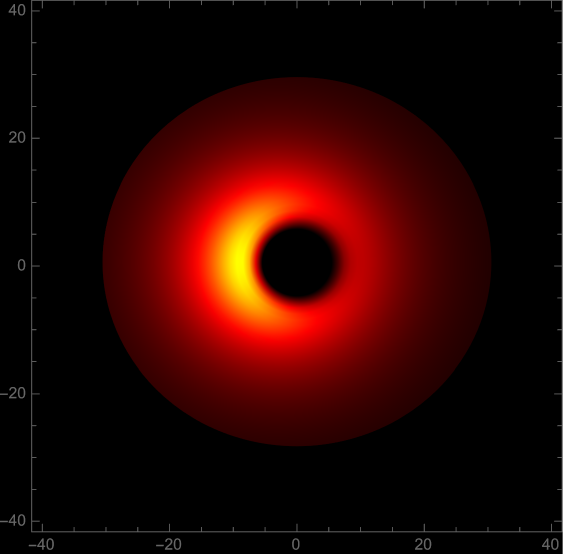}}%
        \usebox{\mybox}%
        \hfill
        \includegraphics[height=\ht\mybox]{color/acc-disk1.pdf}%
        \caption{\scriptsize $a=0.15,\theta_0=20^\circ$}
    \end{subfigure}
    \hfill
    \begin{subfigure}[b]{0.24\textwidth}
        \centering
        \sbox{\mybox}{\includegraphics[width=0.7\linewidth]{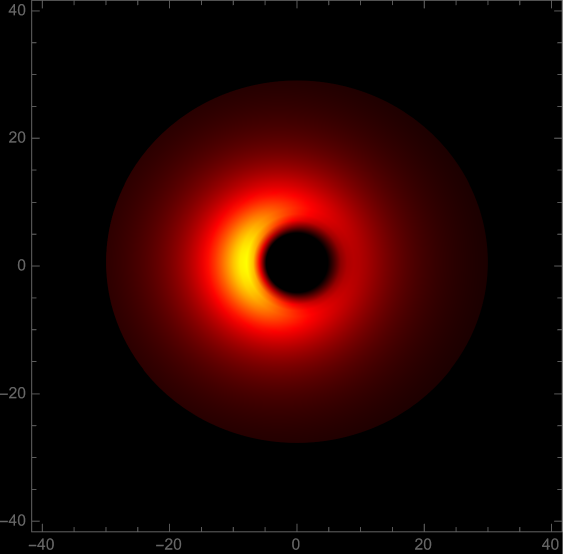}}%
        \usebox{\mybox}%
        \hfill
        \includegraphics[height=\ht\mybox]{color/acc-disk1.pdf}%
        \caption{\scriptsize $a=0.2,\theta_0=20^\circ$}
    \end{subfigure}
    \hfill
    \begin{subfigure}[b]{0.24\textwidth}
        \centering
        \sbox{\mybox}{\includegraphics[width=0.7\linewidth]{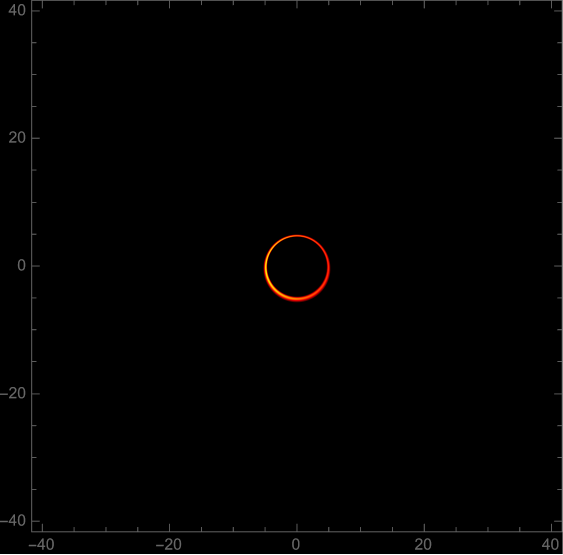}}%
        \usebox{\mybox}%
        \hfill
        \includegraphics[height=\ht\mybox]{color/acc-disk2.pdf}%
        \caption{\scriptsize $a=0.15,\theta_0=20^\circ$}
    \end{subfigure}
    \hfill
    \begin{subfigure}[b]{0.24\textwidth}
        \centering
        \sbox{\mybox}{\includegraphics[width=0.7\linewidth]{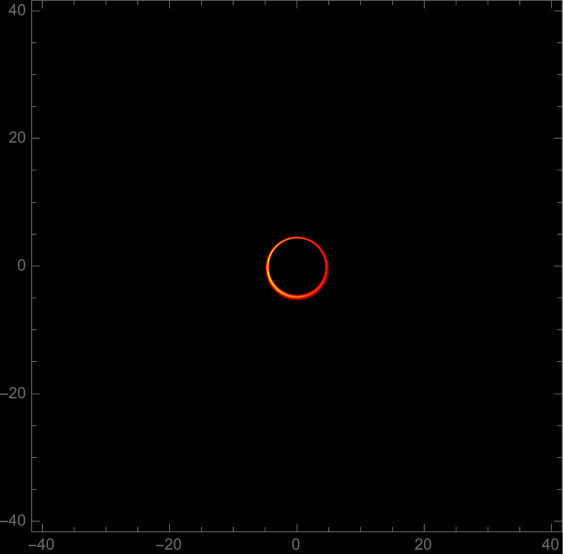}}%
        \usebox{\mybox}%
        \hfill
        \includegraphics[height=\ht\mybox]{color/acc-disk2.pdf}%
        \caption{\scriptsize $a=0.2,\theta_0=20^\circ$}
    \end{subfigure}

    \vspace{0.5cm}

    % Second row (θ₀=60°)
    \begin{subfigure}[b]{0.24\textwidth}
        \centering
        \sbox{\mybox}{\includegraphics[width=0.7\linewidth]{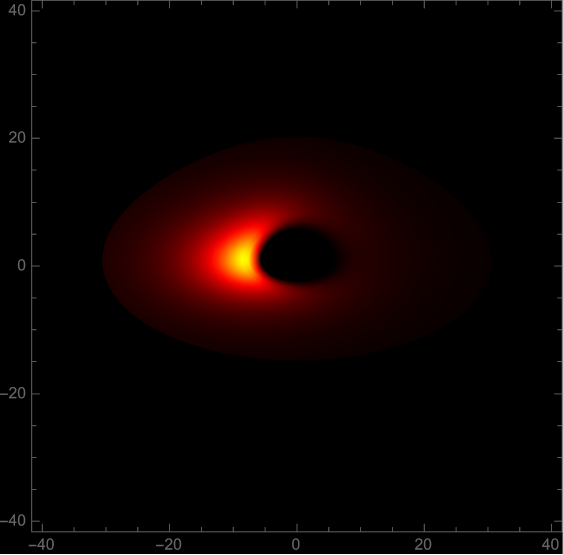}}%
        \usebox{\mybox}%
        \hfill
        \includegraphics[height=\ht\mybox]{color/acc-disk1.pdf}%
        \caption{\scriptsize $a=0.15,\theta_0=60^\circ$}
    \end{subfigure}
    \hfill
    \begin{subfigure}[b]{0.24\textwidth}
        \centering
        \sbox{\mybox}{\includegraphics[width=0.7\linewidth]{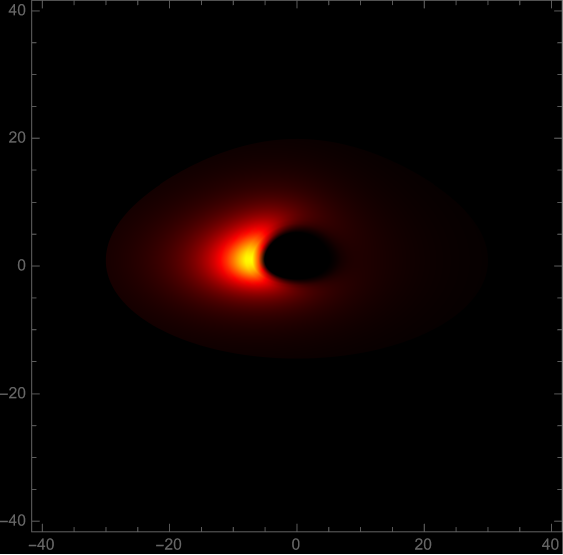}}%
        \usebox{\mybox}%
        \hfill
        \includegraphics[height=\ht\mybox]{color/acc-disk1.pdf}%
        \caption{\scriptsize $a=0.2,\theta_0=60^\circ$}
    \end{subfigure}
    \hfill
    \begin{subfigure}[b]{0.24\textwidth}
        \centering
        \sbox{\mybox}{\includegraphics[width=0.7\linewidth]{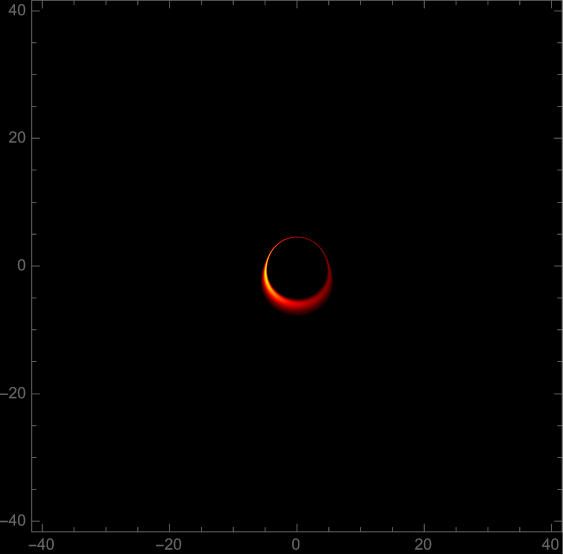}}%
        \usebox{\mybox}%
        \hfill
        \includegraphics[height=\ht\mybox]{color/acc-disk2.pdf}%
        \caption{\scriptsize $a=0.15,\theta_0=60^\circ$}
    \end{subfigure}
    \hfill
    \begin{subfigure}[b]{0.24\textwidth}
        \centering
        \sbox{\mybox}{\includegraphics[width=0.7\linewidth]{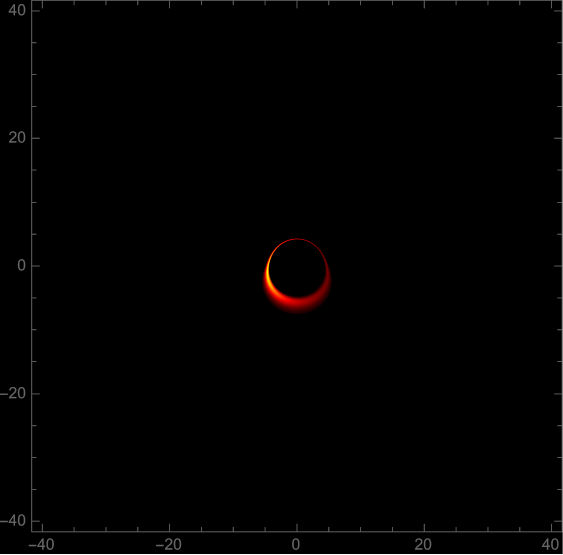}}%
        \usebox{\mybox}%
        \hfill
        \includegraphics[height=\ht\mybox]{color/acc-disk2.pdf}%
        \caption{\scriptsize $a=0.2,\theta_0=60^\circ$}
    \end{subfigure}

    \vspace{0.5cm}

    % Third row (θ₀=85°)
    \begin{subfigure}[b]{0.24\textwidth}
        \centering
        \sbox{\mybox}{\includegraphics[width=0.7\linewidth]{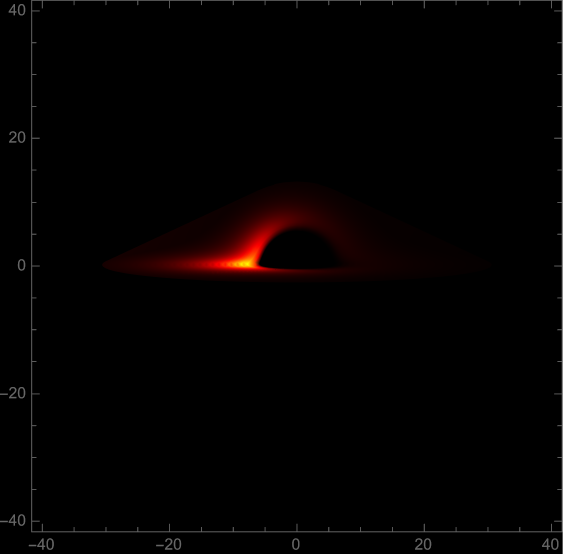}}%
        \usebox{\mybox}%
        \hfill
        \includegraphics[height=\ht\mybox]{color/acc-disk1.pdf}%
        \caption{\scriptsize $a=0.15,\theta_0=85^\circ$}
    \end{subfigure}
    \hfill
    \begin{subfigure}[b]{0.24\textwidth}
        \centering
        \sbox{\mybox}{\includegraphics[width=0.7\linewidth]{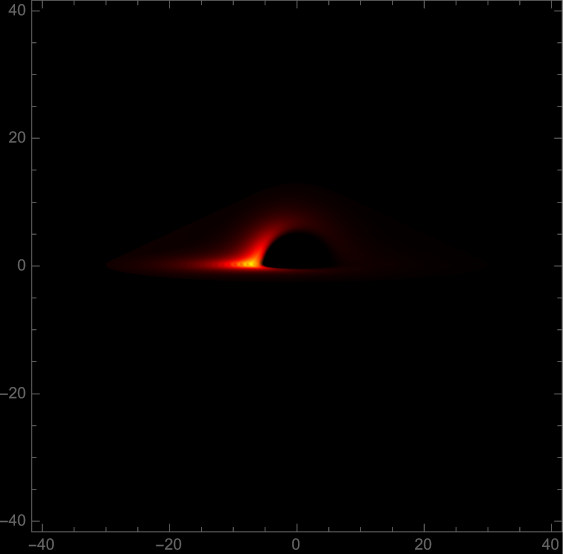}}%
        \usebox{\mybox}%
        \hfill
        \includegraphics[height=\ht\mybox]{color/acc-disk1.pdf}%
        \caption{\scriptsize $a=0.2,\theta_0=85^\circ$}
    \end{subfigure}
    \hfill
    \begin{subfigure}[b]{0.24\textwidth}
        \centering
        \sbox{\mybox}{\includegraphics[width=0.7\linewidth]{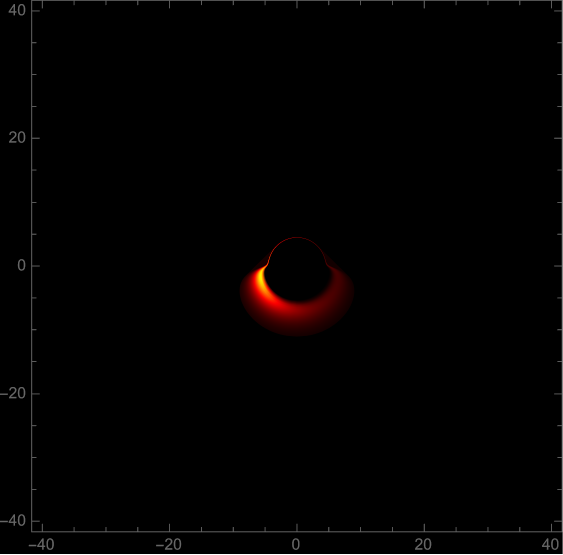}}%
        \usebox{\mybox}%
        \hfill
        \includegraphics[height=\ht\mybox]{color/acc-disk2.pdf}%
        \caption{\scriptsize $a=0.15,\theta_0=85^\circ$}
    \end{subfigure}
    \hfill
    \begin{subfigure}[b]{0.24\textwidth}
        \centering
        \sbox{\mybox}{\includegraphics[width=0.7\linewidth]{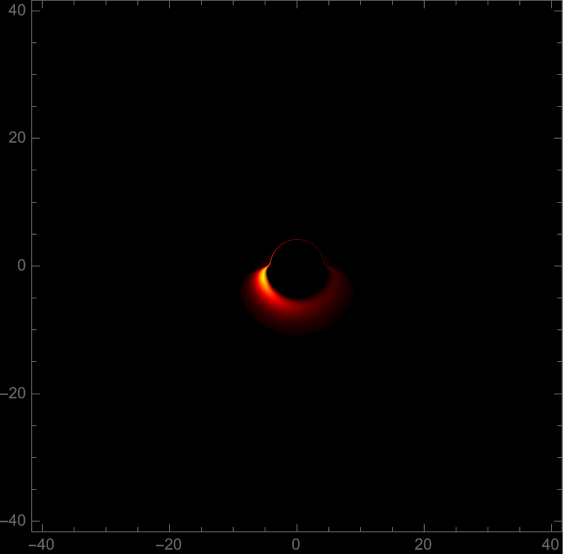}}%
        \usebox{\mybox}%
        \hfill
        \includegraphics[height=\ht\mybox]{color/acc-disk2.pdf}%
        \caption{\scriptsize $a=0.2,\theta_0=85^\circ$}
    \end{subfigure}

    \caption{The influence of the parameter $a$ on the direct and secondary images of the observed radiative flux $F_{\text{obs}}$ at different viewing angles. The outer edge of the accretion disk is at $r=30M$, with $M=1$ fixed.}
    \label{F2}
\end{figure*}

Figs.~\ref{F1} and~\ref{F2} demonstrate the impact of the parameter $a$ on the observed radiation flux $F_{\text{obs}}$ for direct and secondary images at different viewing angles. It can be seen that when the viewing angle is small, the radiation flux distribution of the direct image is approximately disk-symmetric; as the viewing angle increases, the asymmetry of its distribution shape becomes significantly enhanced. Furthermore, an increase in the parameter $a$ leads to a decrease in the ISCO radius, consequently increasing the effective area of the accretion disk. Secondary images correspond to photons that, after being emitted from the disk surface, travel along paths that orbit the black hole by more than $90^\circ$ but less than $180^\circ$ (deflection angle greater than $\pi/2$) before reaching the observer. Such photons are subject to stronger gravitational lensing effects, carrying information from the far side of the disk (relative to the observer). As the viewing inclination increases, the visible region and intensity of the secondary image change.

\section{Conclusion}
\label{Conclusion}

We systematically investigated the shadow, particle circular orbit dynamics, radiation properties of thin accretion disks, and optical imaging characteristics of a static spherically symmetric regular black hole embedded in a Dehnen-type dark matter halo. Quantitative constraints on the model parameter were also provided based on EHT observational data for M87* and Sgr A*. The main conclusions are summarized as follows:

First, by solving the null geodesics, analytical expressions for the black hole shadow radius and critical impact parameter were obtained. Combining these with observational data for the shadow angular diameters of M87* and Sgr A*, we derive the allowed ranges for the model parameter $a$ at the confidence levels $1\sigma$ and $2\sigma$. The results show that the observational constraints from Sgr A* are looser than those from M87*, providing a cross-check opportunity for future multi-messenger constraints.

Second, through a detailed analysis of timelike geodesics, we found that the parameter $a$ has a significant impact on the dynamics of the circular orbit: as $a$ increases, the ISCO radius monotonically decreases, the orbital angular velocity increases, the binding energy of the particles decreases slightly, while the angular momentum shows a clear decreasing trend. Even in the weak-field region far from the horizon, the angular momentum distribution remains sensitive to $a$, offering potential for inferring dark matter halo parameters from accretion disk kinematic features.

Regarding radiation properties, we calculated the local radiative flux, redshift factor, and the observed flux for a distant observer. We employed the backward ray-tracing method to simulate the distribution of isoredshift curves and observed radiative flux in both direct and secondary images. The study shows that an increase in the parameter $a$ reduces the radius of the ISCO and expands the effective radiation area of the accretion disk, thus enhancing the overall observed radiative flux. Analysis of the redshift factor indicates that at small viewing angles, gravitational redshift dominates and exhibits a rotationally symmetric distribution. As the viewing angle increases, the Doppler redshift effect becomes significantly enhanced, the image displays pronounced left-right asymmetry, and the redshift contours are compressed towards one side of the disk. These image features provide intuitive observational criteria for distinguishing between different dark matter halo models and testing black hole metrics.

In summary, this work reveals the systematic modulation patterns of the Dehnen-type dark matter halo parameter on the observational signals of regular black holes, clarifying the intrinsic relationships among the black hole shadow, orbital dynamics, and accretion disk imaging. Future research can be extended in the following directions: (1) Considering the spin effect of the dark matter halo, generalizing the model to axisymmetric rotating cases to explore imaging characteristics and polarization signals in Kerr-like metrics; (2) Incorporating radiation magnetohydrodynamic simulations and accretion flow models with different thicknesses to construct black hole image templates closer to realistic astrophysical environments; (3) Extending the model to include a cosmological constant to study the evolution characteristics of Einstein rings and timelike geodesic structures on cosmological scales; (4) Combining next-generation EHT observations and gravitational wave detectors like LISA to achieve multi-band, multi-messenger joint constraints on dark matter halo parameters. These investigations will help reveal the microscopic physical nature of the interaction between dark matter and black holes in the strong-field regime.

\begin{acknowledgments}
This study is supported in part by National Natural Science Foundation of China (Grant No. 12333008).
\end{acknowledgments}

\FloatBarrier
\bibliographystyle{ieeetr}
\bibliography{reff}
\end{document}